\documentclass{aa}  
\usepackage{graphicx}
\usepackage{txfonts}
\defcitealias{KK20}{I}
\def    \I     {Paper~\citetalias{KK20}}
\begin{document}

   \title{Evaporative cooling of icy interstellar grains}
   \subtitle{II. Key parameters}

   \author{Juris Kalv\=ans
          \and
          Juris Roberts Kalnin
          }
   \institute{
Engineering Research Institute "Ventspils International Radio Astronomy Center" of Ventspils University of Applied Sciences,\\
In$\rm \check{z}$enieru 101, Ventspils, LV-3601, Latvia\\
\email{juris.kalvans@venta.lv}
             }

   \date{Received March 8, 2020; accepted Month DD, YYYY}

  \abstract
   {Evaporative (sublimation) cooling of icy interstellar grains occurs when the grains have been suddenly heated by a cosmic-ray (CR) particle or other process. It results in thermal desorption of icy species, affecting the chemical composition of interstellar clouds.}
   {We investigate details on sublimation cooling, obtaining necessary knowledge before this process is considered in astrochemical models.}
	{We employed a numerical code that describes the sublimation of molecules from an icy grain, layer by layer, also considering a limited diffusion of bulk-ice molecules toward the surface before they sublimate. We studied a grain, suddenly heated to peak temperature $T$, which cools via sublimation and radiation.}
   {A number of questions were answered. The choice of grain heat capacity $C$ has a limited effect on the number of sublimated molecules $N$, if the grain temperature $T>40$\,K. For grains with different sizes, CR-induced desorption is most efficient for rather small grains with a core radius of $a\approx0.02\,\mu$m. CR-induced sublimation of CO$_2$ ice can occur only from small grains if their peak temperature is $T>80$\,K and there is a lack of other volatiles. The presence of H$_2$ molecules on grain surface hastens their cooling and thus significantly reduces $N$ for other sublimated molecules for $T\leq30$\,K. Finally, if there is no diffusion and subsequent sublimation of bulk-ice molecules (i.e., sublimation occurs only from the surface layer), sublimation yields do not exceed 1-2 monolayers and, if $T>50$\,K, $N$ does not increase with increasing $T$.}
   {Important details regarding the sublimation cooling of icy interstellar grains were clarified, which will enable a proper consideration of this process in astrochemical modeling.}

   \maketitle
%

\section{Introduction}
\label{intro}

Thermal desorption of molecules from interstellar grains covered with icy mantles is a process that affects the chemistry in the interstellar medium (ISM) and circumstellar clouds. In an environment with a low ambient temperature, when an icy grain is suddenly heated, such sublimation induces rapid grain cooling. Such a whole-grain heating event may be caused by grain collisions, radioactive decay, protostellar X-rays, or the impact of a heavy cosmic-ray (CR) ion. The latter results in cosmic-ray-induced desorption \citep[CRD,][]{Hasegawa93}.

An initial study that considered the sublimation cooling of interstellar grains in some detail was that of \citet{Herbst06}, who found that icy mantles consisting of a single monolayer (ML) can be completely desorbed in a CRD event. \citet[hereafter, Paper~I]{KK20} investigated the general properties of such sublimation cooling, which happens in competition with radiative cooling of the grain. In that study, we considered a grain with radius $a=0.1\,\mu$m, covered with an icy mantle 100\,MLs thick, rich in volatile molecules. Such or similar grains are expected to reside in interstellar dense, cold ($\approx10$\,K), and dark cloud cores. The main finding of \I\  was that the number of desorbed molecules depends on grain thermal energy, the availability of volatile molecules, and on the fact that grain temperature must exceed a threshold value of about 40\,K for sublimation cooling to dominate over radiative cooling. The desorption yield did not depend on grain cooling time, in opposition to an often-made assumption in papers considering CRD \citep[e.g.,][]{Hasegawa93,Bringa04}.

Our eventual aim is to produce an approach, knowledge, and data on the sublimation cooling of grains in the ISM for applications in astrochemistry. Despite the advances in \I, such an application is not yet straightforward. In the present study, our aim is to remove the main uncertainties related to simulating the sublimation of molecules from grains. The unclear questions are primarily related to the physics and chemistry of icy grains in the ISM. The tasks arise from the aims, as discussed below.

First, we consider the choice of grain heat capacity $C$. A number of authors have employed different approaches for calculating $C$ for different interstellar grain materials, sometimes resulting in conflicting $C$ functions for similar materials. Our task will be to clarify if the choice of $C$ determines the number of molecules thermally desorbed from grains (Sect.~\ref{res-cv}).

Second, grains in the ISM come in a variety of sizes, which affect their absolute heat capacity, surface area, and other properties. We will clarify what the differences are for sublimation for grains with different sizes (Sect.~\ref{res-size}). The need for such a study arises from the uncertainties encountered by studies considering desorption from grains with a variety of sizes, which may result in  grains with different sizes having different ice mantle thicknesses \citep{Herbst06,Pauly16,Iqbal18,Zhao18}.

Third, the composition of the icy mantles on a grain varies for different objects and evolution stages. A few special cases need to be investigated before addressing this problem in future studies. Here we investigate if molecular hydrogen, adsorbed and absorbed in ices, has a role in the cooling of grains and, additionally, if icy grains poor in typical volatiles, such as CO, but rich in CO$_2$ can undergo sublimation cooling as well (Sect.~\ref{res-chem}).

Fourth, there are uncertainties related to molecule diffusion in cryogenic ices. The simulations in \I~considered the diffusion of bulk-ice species, followed by sublimation. Such an inter-layer diffusivity of subsurface molecules is not always considered in astrochemical models, especially those that consider the icy mantle on interstellar grains in a multi-layered manner. To account for such an approach, in Sect.~\ref{res-nd} we investigate the cooling of an icy grain without the diffusion of bulk-ice molecules.

The numerical model for this study is explained below in Sect.~\ref{meth}. The details of the specific tasks and the obtained results are described in Sect.~\ref{resu}. The conclusions are drawn in the final Sect.~\ref{concl}.

\section{Methods}
\label{meth}

We employ the grain cooling model \textsc{Tcool}, presented in \I. The program considers sublimation and radiative cooling of a grain covered with an icy mantle from an initial high temperature $T_0$ to a lower ambient grain temperature, assumed to be $T_2=10$\,K. The initial high temperature $T_0$ depends on the initial thermal energy $E_0$ of the grain with the heat capacity $C$ as the conversion factor between the two. The momentary grain temperature during cooling is $T$. Below we present a concise description of the code. An extended accurate description is presented in \I.

\subsection{Grain model}
\label{grai}

Grains consist of an inert, solid grain core with radius $a$. In \I, the core material was assumed to be olivine. In the present model, grain materials differ only by having a different $C$, which is one of the variables here (Sect.~\ref{res-cv}).
The core is covered with an icy mantle with thickness $b$. Each molecule occupies a volume ${b_m}^3$\,cm$^3$, where $b_m = 3.2\times10^{-8}$\,cm, the assumed size of a molecule, corresponding to water ice with a density of 0.9\,g\,cm$^{-3}$. The molecules in the mantle are arranged in $n$ ice monolayers (MLs). The code treats MLs separately, while molecules of one type in the same ML are treated as arrays. The molecule arrays are chosen, based first on their chemical species and, second, whether they are exposed or not to the outer surface of the grain. Only whole MLs were considered in our previous study; here we employ an updated code that allows a gradual partial depletion of MLs, eliminating some artificial irregularities in the simulation results. A separate array is maintained for the numbers of sublimated, now gas-phase, species.

The default ice composition was described with a monolayer resolution using Eqs.~(18-23) of \I, corresponding to a modeled average ice composition in dark cloud cores. Five potentially volatile molecules were considered -- N$_2$, O$_2$, CO, CH$_4$, and CO$_2$. The remainder was assumed to be water H$_2$O, which also forms the ice matrix, determining properties such as rate of diffusion (Sect.~\ref{loss}). The water ice is the most refractory of the species; the model permits its desorption on sufficient temperature scales and timescales (that in practice never occur in the present study). Some modifications (and simulation results) of the default ice composition are considered in Sect.~\ref{res-chem}.

\subsection{Grain thermal energy loss}
\label{loss}

The initial thermal energy of the heated grain can be defined as
   \begin{equation}
   \label{los1}
         E_0 = \int^{T_0}_{T_2}C(T){\rm d}T \,.
   \end{equation}
In the subsequent cooling from $T_0$ to the final temperature $T_2$, $E_0$ will be lost from the grain by direct molecule sublimation from the surface, the diffusion and subsequent sublimation of bulk-ice molecules, and the emission of photons. The number of molecules on the surface evolves according to the key equation of first-order decay,
   \begin{equation}
   \label{los2}
         N_{\Delta t} = N \times {\rm exp}(-\Delta t/t_{\rm evap}) \,,
   \end{equation}
where $N$ is the initial number of surface molecules, $N_{\Delta t}$ is the number after a time interval $\Delta t$, and $t_{\rm evap}$ is the characteristic sublimation time for the molecule in consideration. The simple case, where Eq.~(\ref{los2}) suffices to describe changes in ice (and the number of sublimated molecules) works only for surface layer molecules on the very first step of the simulation. All other cases are self-consistently supplemented in the code by the logic and consequences resulting from the decrease of molecule numbers in the MLs and, thus, exposure of previously bulk-ice species (with an ever increasing depth) to the surface and their subsequent sublimation (Sects. 2.1.1 and 2.1.2 of \I).

All the icy molecules not exposed to the outer surface have the possibility to diffuse to the surface and subsequently sublimate. \textsc{Tcool} describes diffusive desorption, while also allowing  molecule entrapment in the water ice-dominated layers. The rate is calculated with Eq.~(\ref{los2}), where $N$ denotes the number of bulk-ice species in a particular ML and $t_{\rm evap}$ is replaced by the time of diffusion to the surface summed with the time of sublimation, $t_{\rm diff}+t_{\rm evap}$. We did not consider bulk-ice molecule diffusion \textit{per se}, that is, diffusion that does not result in desorption.

The diffusion time of a molecule depends on its distance to the surface. If a molecule is too deep in the ice, it remains trapped. The data of \citet{Fayolle11} allows us to quantify this effect. Following \I,  for a second-ML molecule, immediately below the surface, the bulk-ice binding energy $E_b$ constitutes $1.1E_D$, with $E_D$ being its desorption energy, a known quantity. For a $n$-th layer MLs, the $E_b$ of a species increases gradually, according to
   \begin{equation}
   \label{los3}
   \begin{array}{l}
      E_{b,n} = 1.1E_{D} \frac{c-(n-2)}{c} + 2E_{D,\rm H_2O} \frac{(n-2)}{c} \\
                        E_{b,n} \leq E_{D,{\rm H_2O}} \,,
   \end{array}
   \end{equation}
where $E_{D,{\rm H_2O}}=5700$\,K is the desorption energy of water, and $c$ is a parameter taken to be 410. This approach describes small and volatile molecule diffusion in a water ice matrix in agreement with experimental data and with the reasonable limitation that their diffusion barriers cannot be greater than $E_{D,{\rm H_2O}}$.

A necessary part of the model is radiative cooling. Below $\sim34$\,K, the rate of energy loss by radiation is higher than the energy lost by the sublimation of N$_2$ and O$_2$. The emission of photons also overtakes sublimation in conditions in which the volatile icy species are depleted (\I). The radiated energy was calculated by integrating emission over photon wavelengths $\lambda$ in the range 2\,$\mu$m--1\,cm according to Eq.~(21) of \citet{Cuppen06}. The lower limit of $\lambda$ means that radiative grain cooling from temperatures $\lesssim700$\,K can be described accurately, more than sufficient for studying stochastic grain heating in dense clouds. This approach does not explicitly consider material-specific aspects of infrared emission, such as the vibrational relaxation of water.

\subsection{Simulation}
\label{simu}

The simulation consists of a series of steps. The results of the separate steps are summed, recording the evolution of temperature, and the numbers of sublimated and ice layer molecules. Each step consists of calculating the number of surface molecules sublimated directly and bulk-ice molecules desorbed with the mediation of diffusion. According to Eq.~(25) of \I, the energy carried away by each molecule is
   \begin{equation}
   \label{sim1}
        E_{\rm evap} = E_D + k_BT \,.
   \end{equation}
This means that each sublimating molecule removes more energy from the grain than just its $E_D$ \citep[e.g., as assumed by][]{Herbst06} and that for higher temperatures and molecules with lower $E_D$ this difference is higher.

The program also calculates the energy lost via radiation, $E_{\rm rad}$ (see previous section). The energies $E_{\rm evap}$ and $E_{\rm rad}$ are combined, obtaining the total energy $E_{\rm cool}$ lost by the grain in the current step. The resulting decrease of temperature $\Delta T$ is then obtained by 
   \begin{equation}
   \label{sim2}
        \Delta T = E_{\rm cool}/C(T) \,,
   \end{equation}
where $C(T)$ is the heat capacity of the icy grain at the current step temperature $T$. The number of remaining icy molecules in each ML is updated in each step. The total number of steps per simulation is on the order of $~10^4$ with $T$ decreased by no more than 0.1\,K per step. The temperature curve is highly irregular and we found it impossible to create a working self-consistent approach for choosing the length of the steps during the simulation. This is because regardless of the initial temperature, molecules are desorbed relatively rapidly at the start of each simulation. Volatile species, such as N$_2$, in the surface layer are quickly depleted, afterwards sublimation continues for volatile subsurface molecules and less-volatile surface species, such as CO or CH$_4$. Finally, while the sufficiently shallow and volatile molecules are being depleted, and the temperature continues to decrease, the grain gradually switches to cooling dominated by photon emission. Because of these complexities, the length for the $m$-th integration step was generally calculated according to
   \begin{equation}
   \label{sim3}
        t_{\rm step}(m) = t_{\rm step}(m-1) + c \times t_{\rm step}(m-1) \,,
   \end{equation}
where $c$ typically is in the range $10^{-2}...10^{-4}$ and can be changed with the help of a simple function during the simulation. The length $t_{\rm step}(1)$ of the very first step and the parameter $c$ were adjusted manually for each simulation, taking into account the most volatile molecule available, the chosen $C$ approach, and the size of the grain. In the output, the code keeps track of the evolving grain temperature, number of sublimated molecules of species $i$, $N_{{\rm ev.}i}$, the remaining icy molecules of the different species, and the amount of energy radiated away.  Test calculations showed that a tenfold increase in the number of steps changes the calculated $N_{\rm ev.}$ by no more than 0.2\,\%.

As discussed in \I, the \textsc{Tcool} model does not consider the pressure of the sublimated gas, which may become important in temperatures above 100\,K when CO and N$_2$ sublimation timescales are $\lesssim10^{-10}$\,s. As the icy molecules are transferred to the gas phase, they may create an expanding `cloudlet' around the grain. If the ices are sufficiently rich in volatiles (as in our assumed standard  ice composition), the cloudlet does not expand fast enough for its interaction with the grain to be completely negligible. While the additional gas pressure will delay sublimation of the remaining icy species, this can only lead to changes in grain cooling time, which does not change $N_{\rm ev.}$ (the radiative cooling timescale is longer by orders of magnitude). A potentially more important effect is that part of the gas will thermalize with the grain. Because the desorbed molecules have a temperature $T_{\rm gas}$ that was possessed by the grain at their moment of desorption (Eq.~(\ref{sim1})), $T_{\rm gas}$ will always be higher than or equal to the current grain temperature $T$. As a result, part of the gas thermal energy can be transferred back to the grain and used for sublimating other molecules. Thus, our calculated numbers of sublimated molecules for grains at temperatures $>100$\,K should be treated as minimum values. We estimate that this effect may increase sublimation by no more than a few per cent, even for grains with $T_0\rightarrow300$\,K.

\section{Results}
\label{resu}

The model described above was used for several simulations, in accordance with the tasks of this study. The specific details of these simulations are described in before the obtained results.

\subsection{Sublimation depending on grain heat capacity}
\label{res-cv}
%
\begin{table*}
\caption{Properties for calculation of $C$ for different grain materials.}
\label{tab-cv}
\centering
\begin{tabular}{l c c c c l}
\hline\hline
Material & $C$ approach & $T_D$, K & \=A\tablefootmark{a}, amu & $\rho$\tablefootmark{b}, g\,cm$^{-3}$ & References \\
\hline
graphite & Draine \& Li (DL) & ... & 12 & 2.24 & \citet{Draine01}, \citet{Xie18} \\
silicate & Draine \& Li (DL) & ... & 22 & 3.32 & \citet{Draine01}, \citet{Xie18} \\
olivine & Leger-Zhao (LZ) & ... & ... & ... & \citet{Leger85}, \citet{Zhao18} \\
quartz & Debye (D) & 542 & 20 & 2.6 & \citet{Xie18} \\
amorphous carbon & Debye (D) & 337 & 12 & 1.557 & \citet{Herbst06}, \citet{Wei05} \\
\hline
\end{tabular}
\\
\tablefoottext{a}{Average atomic mass.}
\tablefoottext{b}{Material density.}
\end{table*}
%
   \begin{figure}
    \vspace{-2cm}
    \hspace{-2cm}
    \includegraphics{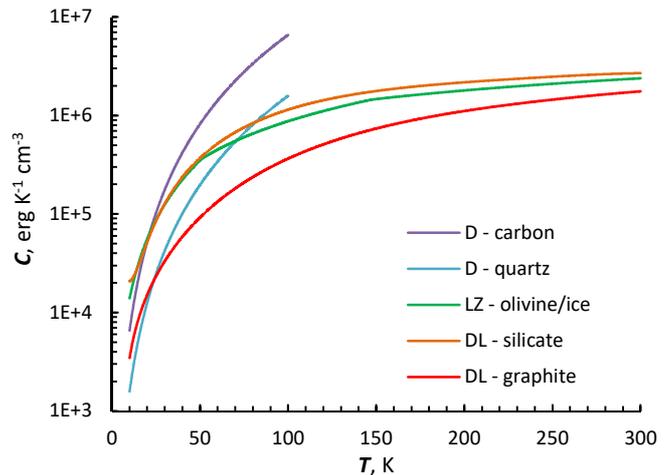}
    \vspace{-22cm}
   \caption{Comparison of grain heat capacities $C$ used in the model. D: The Debye approximation; LZ: Leger-Zhao approach; DL: Draine \& Li approach. Table~\ref{tab-cv} provides details.}
              \label{fig-cv}
    \end{figure}
A crucial parameter in stochastic heating of grains is the heat capacity $C$, which converts grain thermal energy into temperature. The energy a grain receives when hit by a CR particle (or other kinds of energy input) can be calculated directly \citep[e.g.,][]{Shen04}. Conversion to temperature is less straightforward because there are several approaches to calculating $C$ even for similar grain materials. Here we aim to determine, what, if any, effect the choice of $C$ has on the efficiency of molecule sublimation from heated interstellar grains.

A single approach toward $C$ was employed for the whole grain, consisting of a grain core and an icy mantle. Different methods for the $C$ of grain and the $C$ of ice were employed in \I. A single $C$ approach for the whole grain is not entirely physically correct but allows for a simpler reproducibility and clear interpretation of results.

\subsubsection{Calculation of heat capacities}
\label{res-cv-cv}

We employ three different methods for calculating $C$, which, attributed to different materials, result in a total of five approaches. First, a simple method to calculate $C$ is the Debye solid approximation. The heat capacity at a temperature $T\ll T_D$ is
   \begin{equation}
   \label{cv1}
        C = \frac{12\pi^4}{5} N_{\rm at} k_B \left(\frac{T}{T_D}\right)^3 \,,
   \end{equation}
where $N_{\rm at}$ is the number of atoms in the grain. The Debye temperature $T_D$ is in the region of several hundred kelvin. The steep dependence of $C\propto T^3$ means that the Debye approximation is valid only for low temperatures. This condition generally is not fulfilled in the case of CR-induced heating and also photon-induced heating of small grains. However, this approach has been used in astrochemical studies before \citep{Cuppen06,K16}.

In the Debye approximation, different materials primarily differ by $T_D$. A number of values for $T_D$ have been determined for interstellar grains \citep[e.g.,][]{Herbst06}. In this study, we employ two approaches to $C$ with the Debye method, with two extreme values of $T_D$: 337\,K for amorphous carbon \citep{Wei05} and 542\,K for quartz SiO$_2$ \citep{Xie18}.

Analytical equations for $C$ based on experimental data were derived by \citet[their Eq.~(1)]{Leger85} and supplemented by \citet[their Eq.~(13)]{Zhao18}. We adopt the Leger-Zhao method as our second method for calculating $C$. This approach was derived for materials such as olivine and water ice.

The third is a more complex method for $C$, based on a 2D Debye approach by \citet{Draine01} \citep[see also][]{Krumhansl53,Xie18}. Because this approach requires additional integration and is computationally expensive, for practical purposes in the \textsc{Tcool} model, analytical functions of $C$ were derived. The non-dimensional values of $C/(N_{\rm at} k_B)$ can be expressed as
\small
   \begin{multline}
   \label{cv2}
        C/(N_{\rm at} k_B) = -6.91\times10^{-12}T^5 + 6.27\times10^{-9}T^4 - 2.09\times10^{-6}T^3 \\
        + 2.89\times10^{-4}T^2 - 4.79\times10^{-3}T + 3.75\times10^{-2}
   \end{multline}
\normalsize
for silicate and
\small
   \begin{multline}
   \label{cv3}
        C/(N_{\rm at} k_B) = -6.80\times10^{-15}T^6 + 6.65\times10^{-12}T^5 - 2.32\times10^{-9}T^4 \\
        + 2.90\times10^{-7}T^3 + 8.70\times10^{-6}T^2 + 3.25\times10^{-4}T - 2.16\times10^{-3}
   \end{multline}
\normalsize
for graphite. Parameter $N_{\rm at}$ is the total number of atoms in the grain.

Table~\ref{tab-cv} summarizes the properties of the different materials relevant for calculating $C$. When calculating the heat capacity of the icy mantle with the Debye and \citeauthor{Draine01} approaches, the number of atoms in the ice layer (changing with time because of sublimation) was calculated directly from the ice description in the \textsc{Tcool} model (Sect.~\ref{grai}).

Figure~\ref{fig-cv} shows that the values of $C$ calculated with the Debye approximation deviate strongly from those of other methods. The Debye method should not be employed for temperatures exceeding 30--40\,K. However, since our aim is to investigate how variations in $C$ affect sublimation from grains, we include the Debye method in our investigation for $T\leq100$\,K.

\subsubsection{ Sublimation from grains with different $C$}
\label{res-cv-res}

%
   \begin{figure}
 		\vspace{-2cm}
		\hspace{-2.5cm}
   \includegraphics{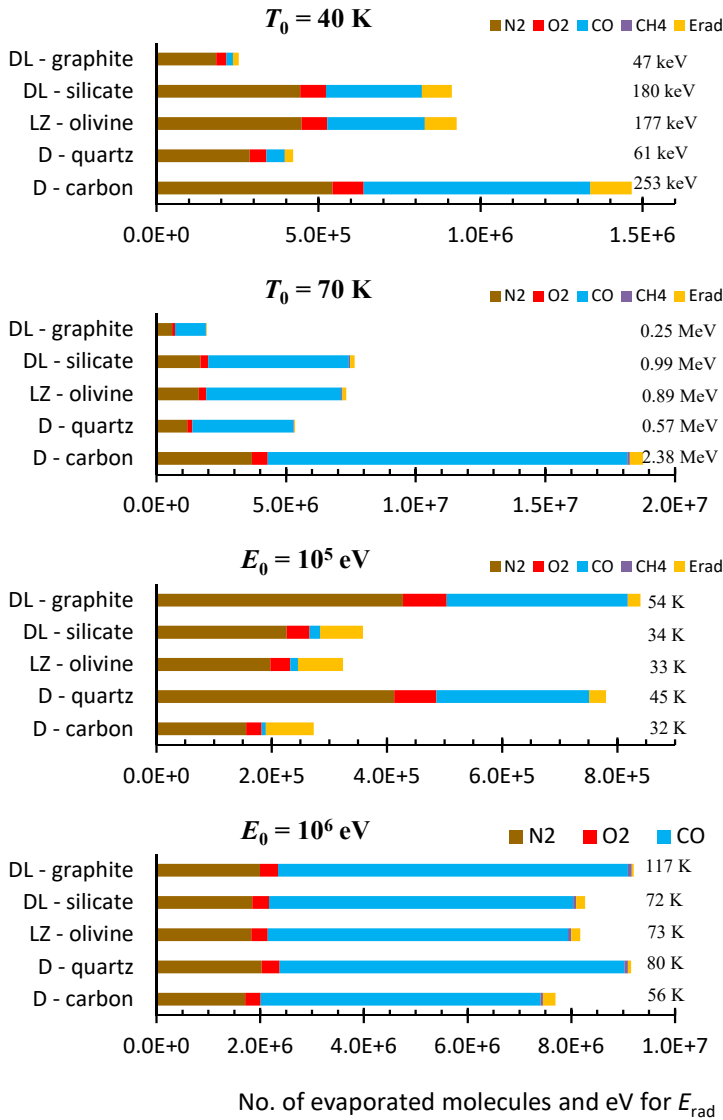}
		\vspace{-13cm}
   \caption{Numbers of different sublimated molecules $N_{\rm ev.}$ for grains with different heat capacities. The abbreviations for heat capacity methods are as in Fig.~\ref{fig-cv}. For simulations with fixed $T_0$, grain initial thermal energies are indicated. For simulations with fixed $E_0$, grain initial temperatures are indicated.}
              \label{fig-cvres}
    \end{figure}
In order to determine the dependence of $N_{\rm ev.}$ on grain heat capacities, simulations with each of the five $C$ variants were performed with two fixed starting temperatures $T_0$. The $T_0$ values were chosen to be 40\,K, which corresponds to a (CR-induced) low heating temperature regime with much of the grain energy lost via radiation, and 70\,K, where grain cooling can be expected to be dominated by thermal desorption.

In addition to fixed $T_0$, we also performed simulations with two fixed grain thermal energies $E_0$. These were chosen to be 0.1\,MeV and 1\,MeV, with considerations similar to those in the choice of $T_0$.

Figure~\ref{fig-cvres} graphically shows that higher $C$ increases the amount of sublimated ices for heating with an equal initial grain temperature $T_0$. The total number of all sublimated molecules per unit energy (sublimation efficiency) is fairly similar for all $C$ approaches, being 4.5--6.5 molecules\,eV$^{-1}$ at 40\,K and 7.5--9.3 molecules\,eV$^{-1}$ at 70\,K. The grain energies $E_0$ differ by almost ten times and the total number of sublimated molecules per grain for different $C$ models differs accordingly, as seen in Fig.~\ref{fig-cvres}.

In the case of $T_0=40$\,K, the CO sublimation efficiency varies more significantly, from 0.4\,eV$^{-1}$ (graphite, \citeauthor{Draine01}) to 2.8\,eV$^{-1}$ (amorphous carbon, Debye) because a larger proportion of N$_2$ and O$_2$ sublimate before CO, cooling the grain if it has a lower $C$ function and accordingly lower $E_0$ (compare Figs.~\ref{fig-cv} and \ref{fig-cvres}). For $T_0=70$\,K, CO is sublimated with an efficiency of 4.7--6.8 CO molecules\,eV$^{-1}$. Models with lower $C$ have a higher desorption efficiency because the total number of sublimated molecules is lower and a higher proportion is accounted for by outer surface molecules that are rapidly removed. Desorption of CO is the CRD process that induces the most profound changes in interstellar cloud chemistry \citep{KK19}. 

A wholly different picture is obtained when we assume a constant initial thermal energy $E_0$ for all grain heat capacity regimes. Materials with lower $C$ functions now have higher temperatures, allowing the sublimation of surface species and efficient diffusion and the subsequent sublimation of bulk-ice species. Especially steep changes are seen if the temperatures $T_0$  are in the vicinity of the sublimation threshold: for $T_0<35$\,K radiative cooling dominates, while for $T_0>40$\,K, most of the thermal energy is carried away by sublimation (see Fig.~\ref{fig-cvres}). At $E_0=0.1$\,MeV, the corresponding sublimation efficiency (for all species) is only around $\sim2$ molecules\,eV$^{-1}$ for simulations with DL-graphite, LZ-olivine, and D-quartz heat capacity methods, while the sublimation efficiency is $\sim8$ molecules\,eV$^{-1}$ for the DL-graphite and D-quartz $C$ methods (for CO these numbers are $\sim0.1$ and $\sim3$ CO molecules\,eV$^{-1}$). For $E_0=1$\,MeV, all grains exceed the sublimation threshold and the sublimation efficiencies are much more similar at 7.5--9.2 molecules\,eV$^{-1}$ and 5.4--6.8 CO molecules\,eV$^{-1}$. Higher sublimation efficiencies are for the simulations with lower $C$ because of their higher initial temperatures.

\subsection{Sublimation depending on grain size}
\label{res-size}
%
\begin{table*}
\caption{Energy and temperature for olivine grains covered with 40\,MLs (0.013\,$\mu$m) icy mantles and hit by CRs depositing the indicated three values of d$E$/d$l$. Data indicating the relative efficiency of CRD for the different grain sizes is also shown.}
\label{tab-size}
\centering
\begin{tabular}{l l c c c c c c c}
\hline\hline
 & Grain core & & & Sublimated & & & $N_{\rm ev.CO}/E_0$\tablefootmark{d}, & \\
No. & size, $\mu$m & $E_0$, eV & $T_0$, K & CO, \%\tablefootmark{a} & $t_{\rm CR hit}$, s\tablefootmark{b} & $R_{\rm des.CO}$\tablefootmark{c} & eV$^{-1}$ & End MLs\tablefootmark{e} \\
\hline
\multicolumn{9}{c}{d$E$/d$l = 2\times10^5$\,eV\,$\mu$m$^{-1}$} \\
1 & 0.2 & 4.1E+04 & 16.5 & 0.0 & 6.1E+09 & 0.0 & 0.0 & 40.0 \\
2 & 0.1 & 2.1E+04 & 23.6 & 0.0 & 4.2E+10 & 0.0 & 0.0 & 40.0 \\
3 & 0.05 & 1.1E+04 & 33.8 & 0.043 & 1.8E+11 & 1.9 & 0.17 & 39.9 \\
4 & 0.02 & 5.3E+03 & 49.9 & 1.2 & ... & 45 & 2.97 & 39.7 \\
5 & 0.01 & 3.3E+03 & 61.8 & 2.0 & ... & 53 & 4.38 & 39.5 \\
\hline
\multicolumn{9}{c}{d$E$/d$l = 10^6$\,eV\,$\mu$m$^{-1}$} \\
6 & 0.2 & 2.1E+05 & 26.5 & 0.0 & 3.9E+10 & 0.0 & 0.0 & 40.0 \\
7 & 0.1 & 1.1E+05 & 39.6 & 1.2 & 2.6E+11 & 11 & 1.64 & 39.7 \\
8 & 0.05 & 5.6E+04 & 57.6 & 6.9 & 9.4E+11 & 58 & 5.31 & 39.0 \\
9 & 0.02 & 2.6E+04 & 91.0 & 16.6 & ... & 100 & 6.60 & 38.2 \\
10 & 0.01 & 1.6E+04 & 117.3 & 24.1 & ... & 90 & 6.64 & 37.7 \\
\hline
\multicolumn{9}{c}{d$E$/d$l = 5\times10^6$\,eV\,$\mu$m$^{-1}$} \\
11 & 0.2 & 1.0E+06 & 44.7 & 7.0 & 4.2E+11 & 7.1 & 3.68 & 39.0 \\
12 & 0.1 & 5.3E+05 & 68.9 & 21.3 & 3.5E+12 & 16 & 5.99 & 37.2 \\
13 & 0.05 & 2.8E+05 & 108.2 & 42.3 & 1.1E+13 & 31 & 6.51 & 34.9 \\
14 & 0.02 & 1.3E+05 & 180.8 & 78.3 & ... & 39 & 6.23 & 31.6 \\
15 & 0.01 & 8.2E+04 & 245.6 & 97.6 & ... & 31 & 5.43 & 30.0 \\
\hline
\hline
 & & $\sigma,\,\mu\rm m^2$ & $N$\tablefootmark{f} & & & \\
\hline
 & 0.2 & 0.13 & 2.1E+8 & & & \\
 & 0.1 & 0.031 & 5.6E+7 & & & \\
 & 0.05 & 0.0079 & 1.6E+7 & & & \\
 & 0.02 & 0.0013 & 3.5E+6 & & & \\
 & 0.01 & 0.00031 & 1.4E+6 & & & \\
\hline
\end{tabular}
\\
\tablefoottext{a}{Percentage of sublimated CO molecules relative to total icy CO; see also Fig.~\ref{fig-sizE}.}
\tablefoottext{b}{Time between CR hits delivering the energy $E_0$ at $A_V=11$\,mag; estimated from \citet{K18}.}
\tablefoottext{c}{Time-averaged desorption rate of the CO molecule; arbitrary units.}
\tablefoottext{d}{Number of CO molecules sublimated per unit of grain thermal energy, eV$^{-1}$.}
\tablefoottext{e}{Final ice thickness in MLs.}
\tablefoottext{f}{Total number of all icy molecules on a grain with a 40\,ML icy mantle.}
\end{table*}

All our studies so far, including \I, have considered grains with a radius of $a=0.1\,\mu$m. However, grains in the ISM are distributed across a variety of sizes and it is crucial to understand the differences of sublimation from grains with different sizes \citep{Herbst06}. This need has been illustrated by the number of assumptions used in the two astrochemical studies focusing on CRD from grains, those of \citet{Iqbal18} and \citet{Zhao18}. Given the lack of understanding and data on sublimation cooling, these papers combine the rate of CR-induced heating of grains with the simple approach on a constant grain cooling time \citep{Hasegawa93} to obtain a method for attributing CRD to large and small grains. Importantly, \citet{Zhao18} scale the cooling time with grain size, while \citet{Iqbal18} do not. Consequently, the former find a higher importance for desorption from large grains, while the latter find that CRD from small grains dominates. In the light of the results from \I, neither of their employed methods are physically rigorous.

\I\  established that sublimation efficiency primarily depends on the heat energy content of the grain \citep[as discussed also by][]{Zhao18}, not its particular temperature and cooling time (given that the threshold temperature of $\sim40$\,K is reached). Here we aim to supplement this qualitative finding by clarifying the differences in the cooling of large and small grains, and also  specifically investigating the case of CR-heated grains.

\subsubsection{Models of grains with different sizes}
\label{res-siz-met}

Grains with sizes of 0.01, 0.02, 0.05, 0.1, and 0.2\,$\mu$m were considered. An equal ice thickness of $n=40$\,ML ($b=0.013\,\mu$m) was assumed for all grains, regardless of size. A similar uniform ice thickness can be expected from the undisturbed accretion of interstellar molecules onto grain surfaces. This means that, while the 0.2+0.013\,$\mu$m grains include 17\,\% ice, the smallest 0.01+0.013\,$\mu$m grains consist of 92\,\% ice by volume. The number of molecules $N$ in the mantles of grains with different sizes vary by two orders of magnitude, as listed in the bottom part of Table~\ref{tab-size}.

The proportions of different icy species differ by a few per cent for different grain sizes. This is because, for smaller grains, the number of molecules in outer MLs is relatively higher than that in the inner MLs close to the grain core. The most significant such difference is for CO, which, along with the less-important O$_2$, is concentrated mainly in the outer MLs close to the surface. For the 0.01\,$\mu$m grain, the $4.58\times10^5$ CO molecules constitute 32.4\,\% of all icy molecules, while for the large 0.2\,$\mu$m grains these numbers are $5.43\times10^7$ and 25.9\,\%, respectively. Such a difference in overall ice composition for grains of different sizes is astrophysically justified, if we adopt the reasonable assumption that grains of all sizes adsorbed a similar chemical mixture from the gas at any given point in cloud evolution.

The heat capacity was adopted from \citet{Leger85} and \citet{Zhao18} for both the olivine core and the icy mantle. The grain starting temperature $T_0$ was chosen based on two complementary approaches. First, we considered an equal temperature for grains of all sizes. Three $T_0$ values were used: 40, 70\,K, and 120\,K. Second, we considered the heating (and subsequent cooling) of grains hit by a single CR-type particle. The property of importance here for such a CR particle is the energy deposited per unit length of the traversed grain material, d$E$/d$l$, which is the stopping power of the fast ion. The energy absorbed by the grain is
   \begin{equation}
   \label{siz1}
   E_0 = {\rm d}E/{\rm d}l \times l\,,
   \end{equation}
where $l$ is the effective path length of a CR traversing the grain. Here we consider CRs that pass through the olivine grain core and ice on both sides of the core. Cosmic rays that only pass through the ice layer were not considered. With the help of the SRIM program \citep{Ziegler10} we found that water ice with an admixture of carbon oxides absorbs about two times less energy than olivine from energetic particles. Thus, we estimate the effective CR path length in the grain as
   \begin{equation}
   \label{siz2}
   l = a + 0.5b.
   \end{equation}
We employed three characteristic values for d$E$/d$l$ for CRs that are able to heat interstellar grains. These are d$E$/d$l = 0.2$\,MeV\,$\mu$m$^{-1}$, 1\,MeV\,$\mu$m$^{-1}$, and 5\,MeV\,$\mu$m$^{-1}$. The first of these values is reached and exceeded, for example, by fast helium nuclei traversing olivine with the $\alpha$-particle energy between 0.07 and 110\,MeV, the second by oxygen nuclei with energies in the range 0.8--100\,MeV, while the third is exceeded by iron nuclei with particle energies of 9--800\,MeV. Table~\ref{tab-size} details the exact energy and temperature reached by grains hit by CRs delivering the indicated d$E$/d$l$.

\subsubsection{Sublimation from grains with different sizes and equal $T_0$}
\label{res-siz-T}

%
   \begin{figure*}
   \centering
		\vspace{-2cm}
    \includegraphics{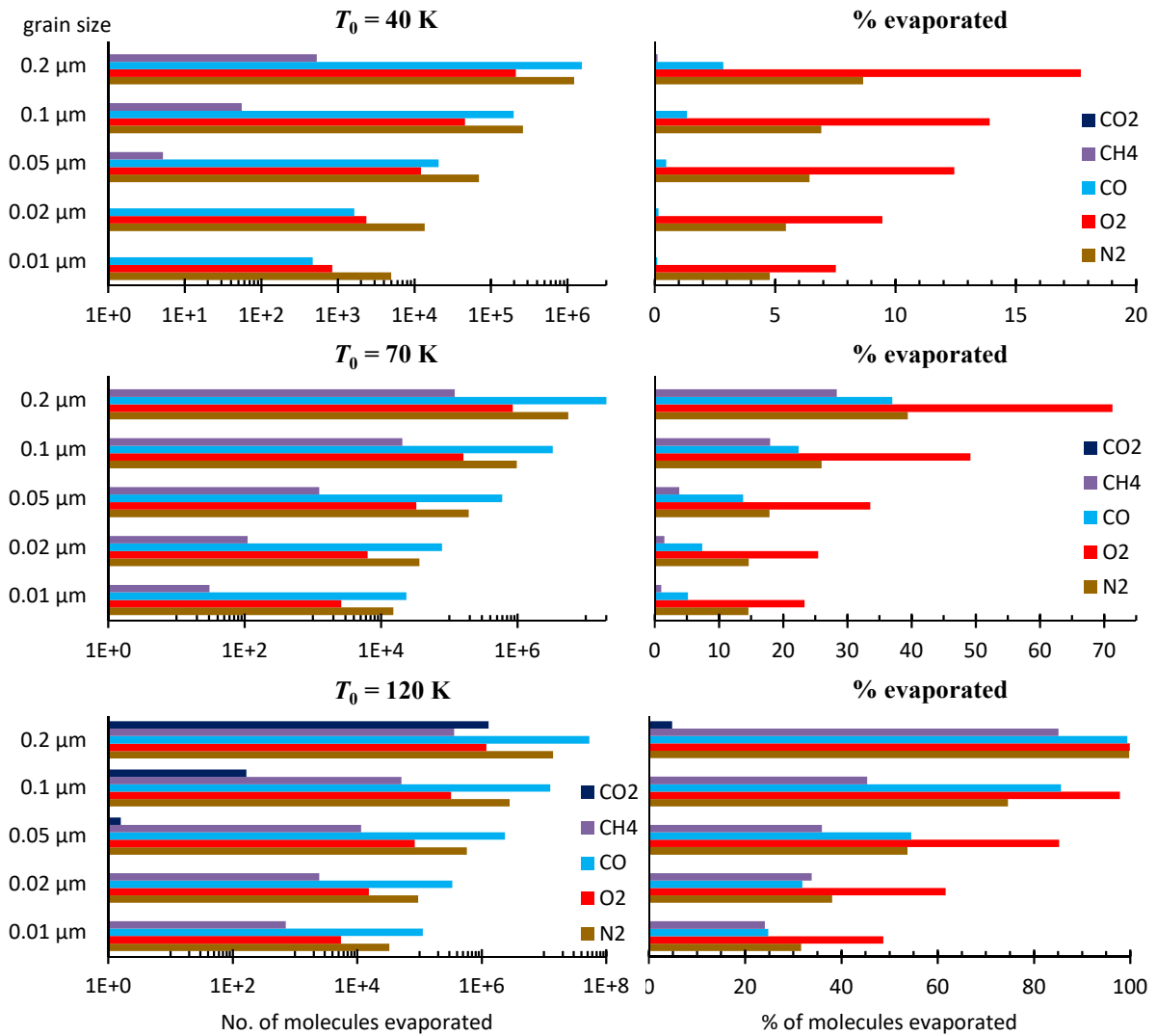}
		\vspace{-14cm}
   \caption{Sublimation of molecules from grains with different sizes, covered with 0.013\,$\mu$m of ice. We show the simulation results for grains with an equal initial temperature $T_0$. Left-hand plots depict the numbers of sublimated molecules $N_{\rm ev.}$, while right-hand plots show the percentage of desorbed molecules relative to the total (initial) number of these molecules.}
              \label{fig-sizT}
    \end{figure*}
Figure~\ref{fig-sizT} shows the number and percentage of various molecules sublimated from grains at 40\,K, 70\,K, and 120\,K initial temperatures $T_0$. As expected, $N_{\rm ev.}$ grows with increasing grain size and higher temperature. Because smaller grains have a larger pool of volatiles relative to their size and contained thermal energy, the small grains may suffice with cooling by N$_2$ and O$_2$ sublimation, with a rather low quantity of sublimated CO ($N_{\rm ev.CO}$). This effect shows for simulations with $T_0=40$\,K and $a\leq0.1\,\mu$m and with $T_0=70$\,K and $a=0.01\,\mu$m.

Several of the simulations result in almost complete desorption for a few species. Molecular oxygen is most easily depleted because it is the most volatile species (along with N$_2$), is concentrated near the surface, and has a rather low overall abundance. The latter aspect means that the sublimation of O$_2$ ice cannot appreciably cool the grain and thus O$_2$ cannot prevent the depletion of itself. Consequently, O$_2$ sublimation percentages are always higher than those of other species and approach 100\,\% in simulations considering large grains and high temperatures.

The two simulations with $T_0=120$\,K and $a\geq0.1\,\mu$m result in the near-complete depletion of volatiles, with the percentage of desorbed O$_2$, N$_2$, and CO exceeding 75\,\%. For the 0.2\,$\mu$m grain, more than 99\,\% of the molecules of these three species are depleted from ices, while 85\,\% of CH$_4$  also being sublimated. This is the only simulation that shows significant sublimation of CO$_2$ at 4.9\,\% level. Ninety percent of the CO$_2$ molecules sublimate between temperatures of 90\,K and 80\,K.

The high level of sublimation from the 120\,K and 0.2\,$\mu$m grain occurs because it contains the highest amount of thermal energy of all 15 grains considered in this subset of simulations. Moreover, it has the least number of icy molecules ($2.1\times10^8$) versus thermal energy ($E_0=14.1$\,MeV). This is because the number of molecules approximately depends on the surface area of the grain, while its heat capacity and, thus $E_0$, depends on its volume.

\subsubsection{Sublimation from grains with different sizes hit by CRs}
\label{res-siz-E}

%
   \begin{figure*}
   \centering
		\vspace{-2cm}
    \includegraphics{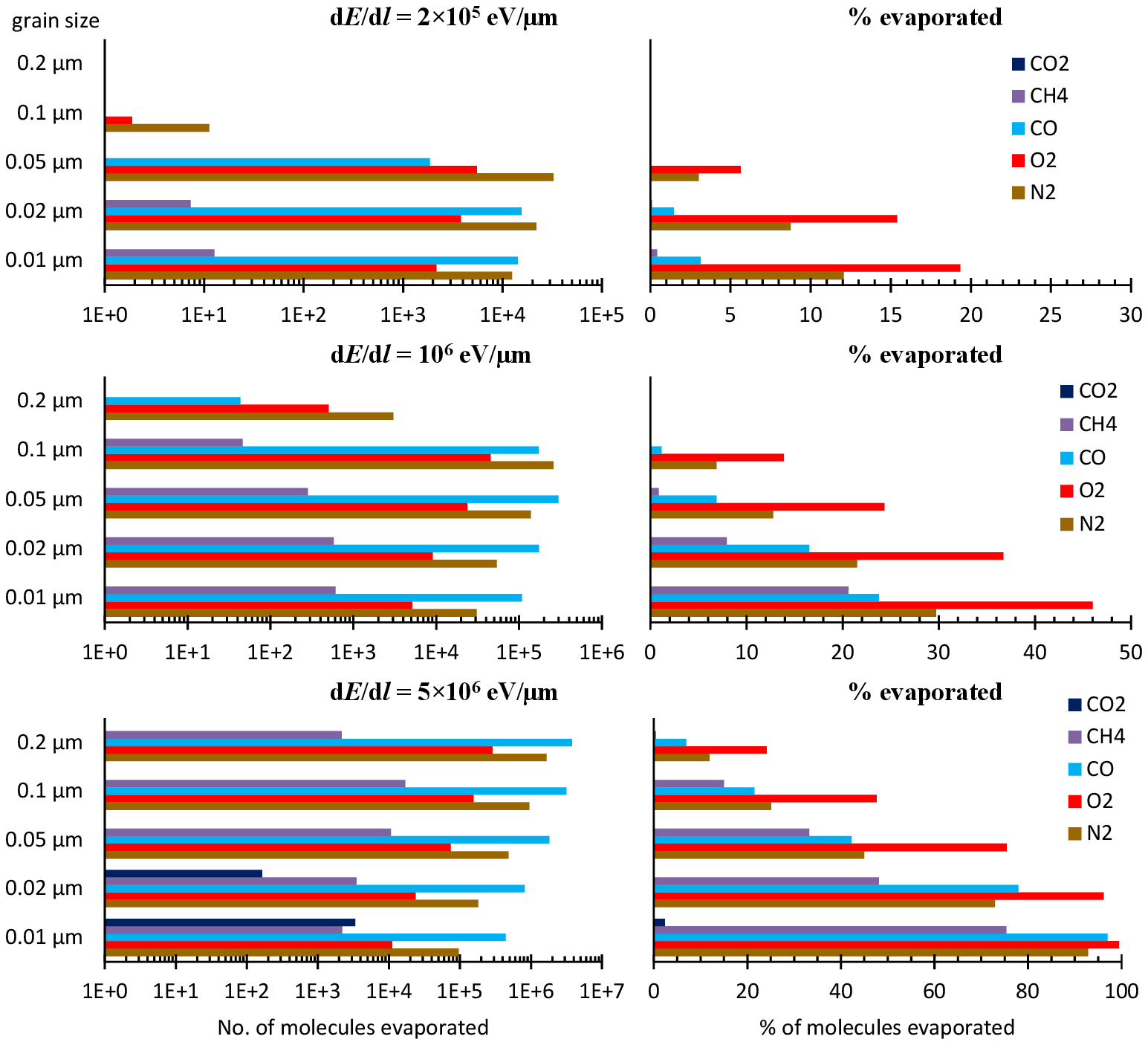}
		\vspace{-14cm}
   \caption{Sublimation of molecules from grains with different sizes, covered with 0.013\,$\mu$m of ice. We show the simulation results for grains that were hit by CRs with three different stopping power d$E$/d$l$ values. Left-hand plots depict the numbers of sublimated molecules $N_{\rm ev.}$, while right-hand plots show the percentage of desorbed molecules relative to the total (initial) number of these molecules.}
              \label{fig-sizE}
    \end{figure*}
The above section treated the cooling of icy grains as a purely theoretical phenomenon. In this section, we aim to explore the same process as directly initiated by CR-induced heating. To do so, the initial temperatures of the grains with different sizes were calculated after they had been hit by CR particles, as explained in Sect.~\ref{res-siz-met}. Hits by three types of CR particles were considered. Their stopping powers d$E$/d$l$ and the calculated $N_{\rm ev.}$ are shown in Table~\ref{tab-size}. The first two columns of this table show the exact thermal energy $E_0$ deposited by the CRs according to Eq.~\ref{siz1} and the initial temperature $T_0$ reached by the grains, according to Eq.~(\ref{los1}), before the onset of cooling. Because the effective CR path length $l$ in a grain depends on the grain radius, while its heat capacity is proportional to radius to the third power, the smaller grains are heated to much higher temperatures than the large grains, when hit by the same type of CR particles.

The top plots of Fig.~\ref{fig-sizE} show the sublimation from grains with $T_0$ in the range 16--62\,K. This range crosses the 30--40\,K threshold. For temperatures below this threshold, radiative cooling is faster than sublimation, while for temperatures above the threshold, cooling can be dominated by the sublimation of N$_2$ and CO if these molecules are present. Therefore, the 0.2\,$\mu$m grains at 16.5\,K show no thermal desorption at all, 0.1\,$\mu$m grains at 34\,K show very limited sublimation, while the smaller grains at higher temperatures are able to sublimate a noticeable part of their icy volatiles.

Aside from the effect of the sublimation threshold temperature, the $N_{\rm ev.}$ values are much more comparable for grains with different sizes than in the case for grains with equal $T_0$ in Sect.~\ref{res-siz-T}. Such a similarity occurs because the thermal energies $E_0$ of the grains are now much more similar. This result underlines the conclusion from \I\  that $N_{\rm ev.}$ primarily depends on $E_0$ and not on the exact $T_0$ or time of cooling. Given that the smaller grains have a smaller pool of volatiles, the percentage of sublimated molecules is significantly higher for small grains. Nevertheless, only 0.02\,$\mu$m and 0.01\,$\mu$m grains, assumed to be hit with the highest energy CRs, approach a near-complete depletion of volatiles from ice, with $>70$\,\% of N$_2$, O$_2$, and CO molecules being desorbed.

The highest temperature grains (181\,K and 246\,K) also show that CO by percentage is sublimated more than N$_2$, which is unlike the results of all other simulations and is unusual because N$_2$ has a lower desorption energy than CO (1000\,K versus 1150\,K). This phenomenon can be explained by the concentration of CO in the upper layers of the icy mantle and its rapid sublimation on a timescale of about $10^{-10}$s. N$_2$ has proportionally more molecules in ice depth, from where it takes more time to diffuse to the surface before sublimation. A large part of such near-surface CO sublimates, quickly lowering $T$ and thus reducing the rate of N$_2$ and other molecule diffusion from deeper layers below.

It is possible to estimate the relative effectiveness of CRD for grains with different sizes by multiplying the percentage of sublimated CO molecules from each grain type by the grain cross section $\sigma$, which is proportional to CR hit rate. These data can be compared between the three CR types considered thanks to the known time $t_{\rm CR}$ between CR-grain collisions depositing the indicated $E$ in 0.2\,$\mu$m, 0.1\,$\mu$m, and 0.05\,$\mu$m grains from \citet{K18}. Finally, by relating the obtained values in an inversely proportional manner to the overall number of icy molecules on a grain of a specific size, we obtain an estimate of how efficient the CR-induced desorption of CO is for grains of different sizes by carrying an equal mass of ice for each grain size bin. Table~\ref{tab-size} details the values of all the mentioned parameters; the relative CO desorption rate $R_{\rm des.CO}$ is given on a scale from 0 to 100.

In this way we find that, of all the sizes considered, CRD from 0.01\,$\mu$m and 0.02\,$\mu$m grains hit by moderate energy CRs is most efficient. Here we assume an equal ice mass distribution among the given sizes of grains. Neither a realistic interstellar grain size distribution, nor a realistic accretion scenario of molecules onto grains was considered; a full calculation of the CRD rate is the task of models involving more interstellar physics.

Finally, the last column of Table~\ref{tab-size} quantifies the dependence of sublimation on grain thermal energy. The highest amount of sublimated CO molecules per unit of energy is $N_{\rm ev.CO}/E_0=6...7$\,eV$^{-1}$, achieved by grains heated to high temperatures and sufficient reserves of icy CO for sublimation. The highest temperature grains suffer from lack of CO and other volatiles for cooling, while low-temperature grains lose a major part of their energy in radiative cooling and in the desorption of N$_2$.

\subsection{Sublimation from grains with mantles of specific chemical composition}
\label{res-chem}

Interstellar ices can possibly have a wide variety of compositions \citep[e.g.,][]{Oberg11}. Our task here is not to explore the whole parameter space but to investigate questions that must be clarified before sublimation cooling can be applied for astrochemical models. We identify two such questions: determining the ability of CO$_2$ ice sublimation and determining if adsorbed H$_2$ can play a significant role in grain cooling.

\subsubsection{Cooling of CO$_2$-rich grains}
\label{res-co2}

%
   \begin{figure}
		\vspace{-2cm}
		\hspace{-2cm}
    \includegraphics{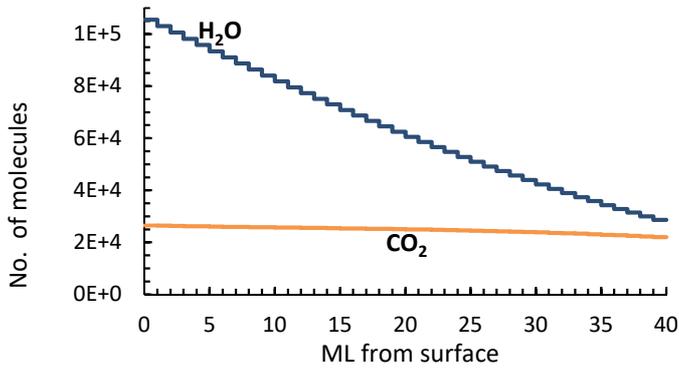}
		\vspace{-23cm}
   \caption{Adopted number of molecules per ML for the CO$_2$-rich mantle for a grain with $a=0.02\,\mu$m olivine core. The total number of molecules per ML is higher towards the surface because the ice layers increase the size of the grain.}
              \label{fig-co2}
    \end{figure}
%
\begin{table*}
\caption{Simulation results for sublimation from small grains, assumed to be heated by CRs with given stopping power d$E$/d$l$, covered with 40\,ML ice consisting of CO$_2$ and H$_2$O.}
\label{tab-co2}
\centering
\begin{tabular}{lcccccc}
\hline\hline
Grain core &  &  &  & \% CO$_2$ & \% $E_0$ & End \\
size, $\mu$m & $E_0$, eV & $T_0$, K & $N\rm_{ev.CO_2}$ & evap.\tablefootmark{a} & evap.\tablefootmark{b} & MLs\tablefootmark{c} \\
\hline
\multicolumn{7}{c}{d$E$/d$l = 10^6$\,eV\,$\rm \mu m^{-1}$} \\
0.05 & 5.6E+04 & 57.6 & 0.0E+00 & 0.0 & 0.0 & 40.0 \\
0.02 & 2.6E+04 & 91.0 & 1.3E+04 & 1.3 & 11.6 & 39.9 \\
0.01 & 1.6E+04 & 117.3 & 2.3E+04 & 6.1 & 32.9 & 39.6 \\
\hline
\multicolumn{7}{c}{d$E$/d$l = 5\times10^6$\,eV\,$\rm \mu m^{-1}$} \\
0.05 & 2.8E+05 & 108.2 & 1.8E+05 & 4.0 & 15.0 & 39.6 \\
0.02 & 1.3E+05 & 180.8 & 3.0E+05 & 30.3 & 53.6 & 37.7 \\
0.01 & 8.2E+04 & 245.6 & 2.2E+05 & 57.8 & 64.4 & 36.4 \\
\hline
\end{tabular}
\\
\tablefoottext{a}{Percentage of sublimated CO$_2$ molecules relative to the total initial number of icy CO$_2$.}
\tablefoottext{b}{Percentage of initial grain thermal energy carried away by CO$_2$.}
\tablefoottext{c}{Final ice thickness in MLs.}
\end{table*}
Observations have shown that a variety of ice chemical compositions are possible in the ISM. Among these are ices consisting of carbon dioxide, water, and, perhaps, an elevated amount of methanol, but with no observed volatiles  like CO and CH$_4$ \citep{Boogert11,Whittet11}. Such ices can arise in the ISM through prolonged or intense photoprocessing of solid CO:H$_2$O mixtures by interstellar UV photons \citep{Woon04,KS10}. In order to obtain a picture on the sublimation of such heated CO$_2$-rich icy grains, the \textsc{Tcool} model was applied for an olivine grain with $C$ calculated with the \citeauthor{Leger85} and \citeauthor{Zhao18} approach, and coated with a 40\,ML icy mantle. Because CO$_2$ sublimation cannot compete with radiative cooling for temperatures lower than $\sim80$\,K, we considered only the smaller grains that reach higher temperatures when hit by CRs. Cooling of $a=0.05\,\mu$m, 0.02\,$\mu$m, and 0.01\,$\mu$m grains was modeled, as they were heated by CRs with stopping powers d$E$/d$l=1$\,MeV and 5\,MeV.

The icy mantle was assumed to consist only of CO$_2$ and H$_2$O. In the ISM, the observed CO$_2$:H$_2$O ratios for ices  with no detection of the volatiles CO or CH$_4$ lie in the range 23...188\,\% with a median value of 38\,\% \citep{Boogert11,Boogert13,Whittet11}. The relative abundances of N$_2$, O$_2$, CO, and CH$_4$ were taken to be 0, while that of CO$_2$ was calculated by adding 17\,\% (relative to the number of all icy molecules) to Eq.~(22) of \I,
\begin{equation}
        \label{codivi1}
        X_{\rm CO_2} = 0.12x^2 + 0.12x + 0.20\,,
\end{equation}
where $X_{\rm CO_2}$ is the proportion of CO$_2$ in an ice ML that is located at a relative depth $x$ in the mantle (i.e., $x$ is expressed as part of the mantle thickness). In this way we obtain CO$_2$:H$_2$O overall ice abundance ratios of 37\,\%, 39\,\%, and 41\,\% for the 0.01\,$\mu$m, 0.02\,$\mu$m, and 0.05\,$\mu$m grains, respectively. Figure~\ref{fig-co2} shows that the number of CO$_2$ molecules is higher in shallow layers, although its relative proportion is higher in the deeper layers. CO$_2$ covers 20\,\% of the icy surface, while its proportion in the inner ML adjacent to the grain core is 43\,\%. The shallow MLs have a higher number of adsorption sites due to the increase of grain size with each ML. Because of a higher $E_D$, each desorbed CO$_2$ molecule carries away about twice as much heat from the grain as the lighter volatiles N$_2$ and CO.

Table~\ref{tab-co2} shows $E_0$, $T_0,$ and the modeling results for the cooling of small grains via sublimation of CO$_2$. The primary finding is a verification that CRD can induce desorption of CO$_2$ albeit only from small grains. CO$_2$ sublimation occurs to a significant extent for grain temperatures of $>100$\,K; an energetic CR impact may result in the removal of more than half of all the CO$_2$ inventory of a grain. Cosmic-ray-induced desorption efficiency for CO$_2$ desorption is limited by the inability of CRs to sufficiently heat medium-sized and large grains. Nevertheless, CRD may serve as a source of gaseous CO$_2$ in interstellar clouds.

\subsubsection{Sublimation from grains with H$_2$ absorbed in ice}
\label{res-h2}
%
   \begin{figure}
		\vspace{-2cm}
		\hspace{-2cm}
    \includegraphics{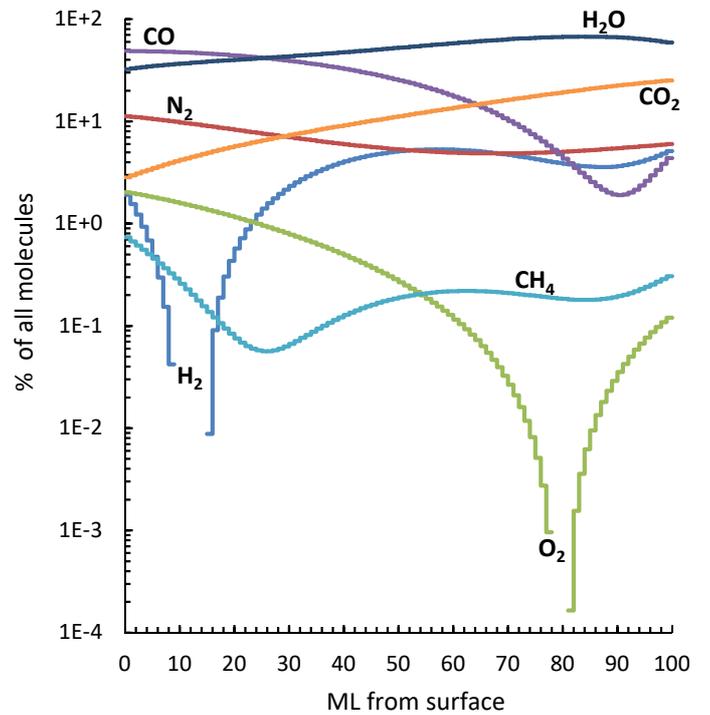}
		\vspace{-18.5cm}
   \caption{Adopted relative abundance per layer (percentage relative to all icy molecules in that layer) of H$_2$ and other molecules for a 100\,ML ice mantle. For simulations not considering hydrogen, the H$_2$ abundance value was added to that of water (cf. Fig.~2 of \I).}
              \label{fig-arH2}
    \end{figure}
%
   \begin{figure}
		\vspace{-2cm}
		\hspace{-1.5cm}
    \includegraphics{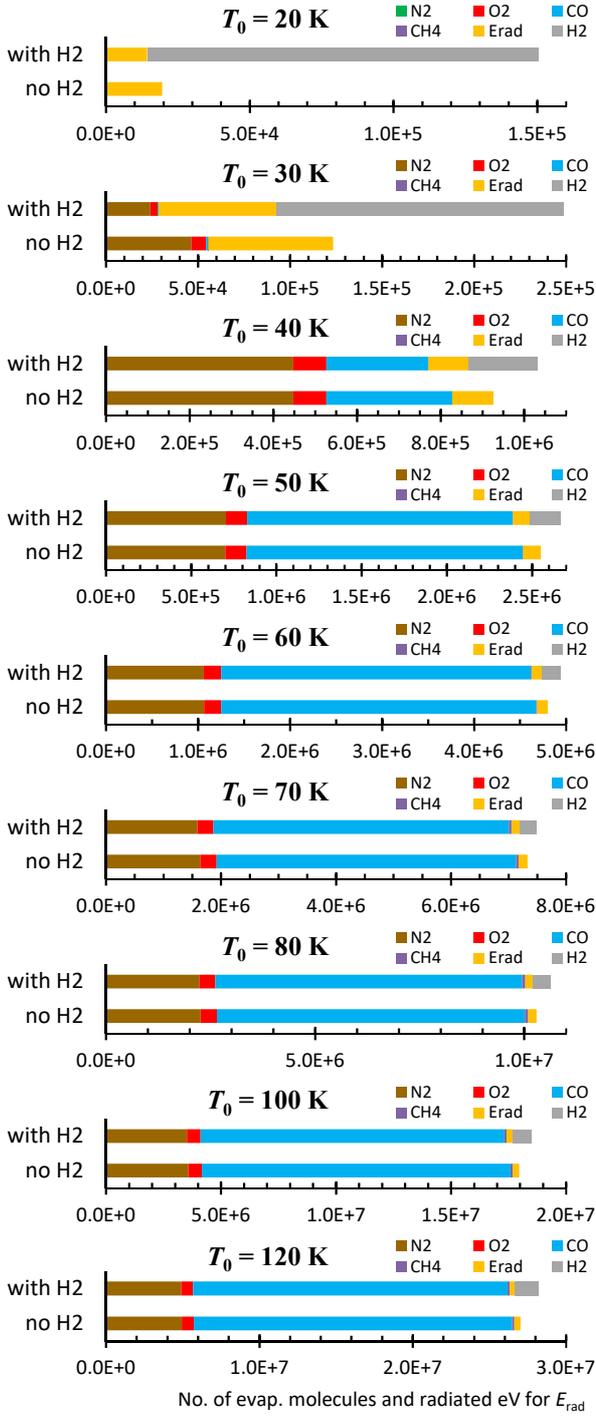}
		\vspace{-9cm}
   \caption{Comparison of results for simulations with and without H$_2$ molecules absorbed in icy mantle -- $N_{\rm ev.}$ for grains with different initial temperatures.}
              \label{fig-h2res}
    \end{figure}
%
\begin{table*}
\caption{The number of sublimated H$_2$ ($N_{\rm ev.H_2}$), the percentage of energy H$_2$ carried away, and the total number of all other sublimated molecules (N$_2$, O$_2$, CO, and CH$_4$) expressed as a percentage relative to simulations without H$_2$.}
\label{tab-h2}
\centering
\begin{tabular}{l c c c}
\hline\hline
Initial & No. of sublimated & \% of energy & Other sublimated molecules \\
$T$,\,K & H$_2$ molecules, $\times10^5$ & carried away & with vs. without H$_2$, \% \\
\hline
20 & 1.36 & 26.9 & ... \\
30 & 1.56 & 8.6 & 51.6 \\
40 & 1.64 & 4.0 & 93.1 \\
50 & 1.77 & 2.2 & 97.5 \\
60 & 2.01 & 1.5 & 98.7 \\
70 & 2.76 & 1.4 & 98.3 \\
80 & 4.12 & 1.5 & 99.3 \\
100 & 8.18 & 1.8 & 98.5 \\
120 & 15.13 & 2.3 & 98.9 \\
\hline
\end{tabular}
\end{table*}
When considering CRD yields, an uncertain role is played by hydrogen adsorbed on the icy surface and absorbed in ice. As the most volatile species, any H$_2$ molecules \citep[and perhaps H atoms,][]{Rawlings13} can be expected to sublimate from a heated grain before other molecules, rapidly cooling the grain. Such desorption would have little effect on the gas-phase abundance of H$_2$ but would rob the grain of thermal energy, which thus cannot be used for sublimating other icy species.

The astrochemical model \textsl{Alchemic-Venta} predicts icy H$_2$ relative abundances on the order of a few per cent, and a negligible abundance for H atoms (see the method for obtaining abundances in \I\  and references therein). The respective abundance function for H$_2$ is
        \begin{multline}
           \label{hdivi1}
   X_{\rm H_2} = 1.77x^4 - 3.73x^3 + 2.44x^2 - 0.451x + 0.0238 \\
         X_{\rm H_2} \geq 0\,.
   \end{multline}
Figure~\ref{fig-arH2} shows that the resulting abundance of H$_2$ absorbed in ice MLs is higher near the surface and in the inner part of the mantle. Elevated H$_2$ abundance in the surface MLs arises because of H$_2$ adsorption from the gas in a molecular cloud. This can be described as an equilibrium adsorption--sublimation process, which means that surface H$_2$ can be quickly replenished between hits of CR particles. The elevated abundance of H$_2$ in the inner MLs, near the inert core of the grain, arises because of the photoprocessing of water-rich ice layers containing carbon monoxide, the same process that generates icy CO$_2$ in the deep layers of the mantle:
\begin{displaymath}
{\rm H_2O + CO} \stackrel{h\nu}{\rightarrow} ... \rightarrow {\rm CO_2 + H_2} \,.
\end{displaymath}
Such processing was experimentally shown to occur in a H$_2$O:CO icy mixture by \citet{Woon04}, while the model of \citet{KS10} showed its relevance for the ISM. Modeling shows that part of the generated H$_2$ may remain absorbed in ice. These H$_2$ reserves cannot be quickly replenished and the timescales for H$_2$ abundance build-up are hundreds of thousands of years.

To model the sublimation of H$_2$-containing ices we employed the \citet{Leger85} heat capacity calculation. Figure~\ref{fig-h2res} shows a comparison of calculation results at different initial grain temperatures: $N_{\rm ev.}$ for volatile molecules for icy mantles with absorbed H$_2$ and without H$_2$.

Table~\ref{tab-h2} shows data characterizing the effects of the addition of adsorbed and absorbed H$_2$ to the icy mixture. The low-$T$ regimes are most severely affected. For increasing $T_0$, until $T=80$\,K, the $N_{\rm ev.H_2}$ remains low, within about a few times $10^5$, and practically only the shallow H$_2$ reservoir is depleted. Higher temperatures are able to induce significant diffusive sublimation of the deeper, photochemically generated H$_2$ reservoir, which shows up in the increase of the thermal energy part carried away by H$_2$.

To summarize, the grain heating regimes can be divided into two classes -- low $T$ regimes that are strongly affected by the addition of the adsorbed H$_2$, and medium and high $T$ regimes, where the effects of adsorbed and absorbed H$_2$ are limited. This result is important because \citet{KK19} found that the low-$T$ CRD regimes can be quite efficient at desorbing volatiles. This finding, according to our new results, seems not to be the case.

\subsection{Mantle sublimation with no bulk-ice diffusion}
\label{res-nd}
%
   \begin{figure}
		\vspace{-2cm}
		\hspace{-1.5cm}
    \includegraphics{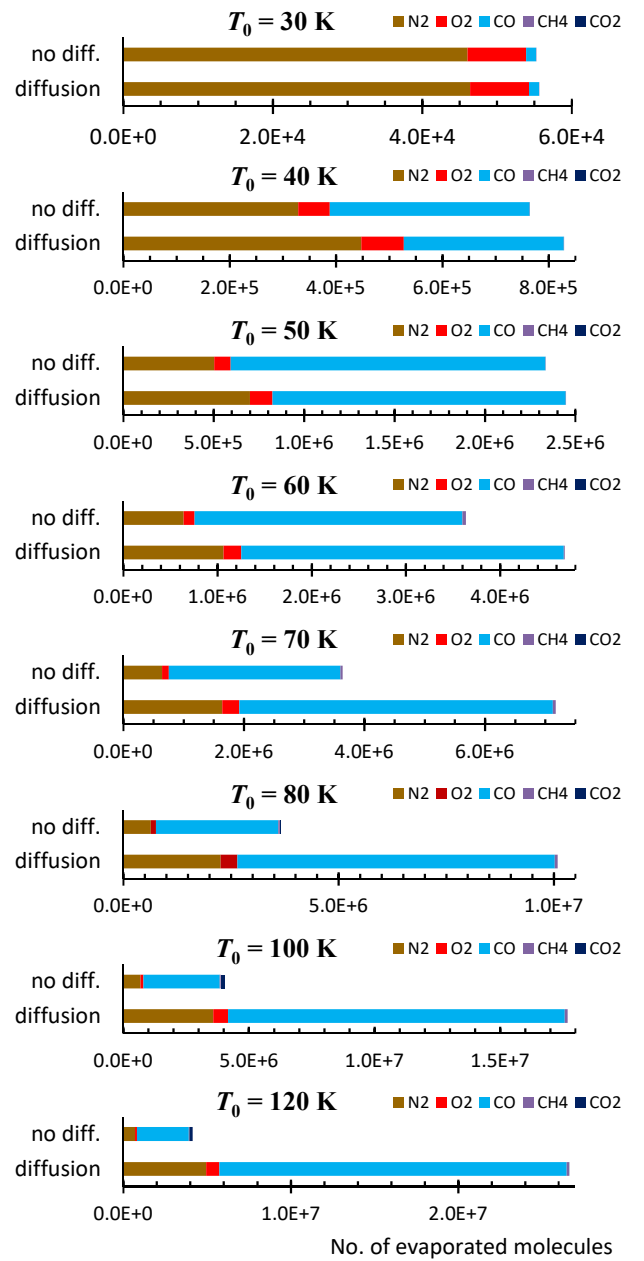}
		\vspace{-11cm}
   \caption{Comparison of results for grain cooling simulations with and without diffusion of molecules allowed from the subsurface bulk-ice layers of the icy mantle -- $N_{\rm ev.}$ for grains with different initial temperatures.}
              \label{fig-nd}
    \end{figure}
%
   \begin{figure*}
   \centering
		\vspace{-2.0cm}
    \includegraphics{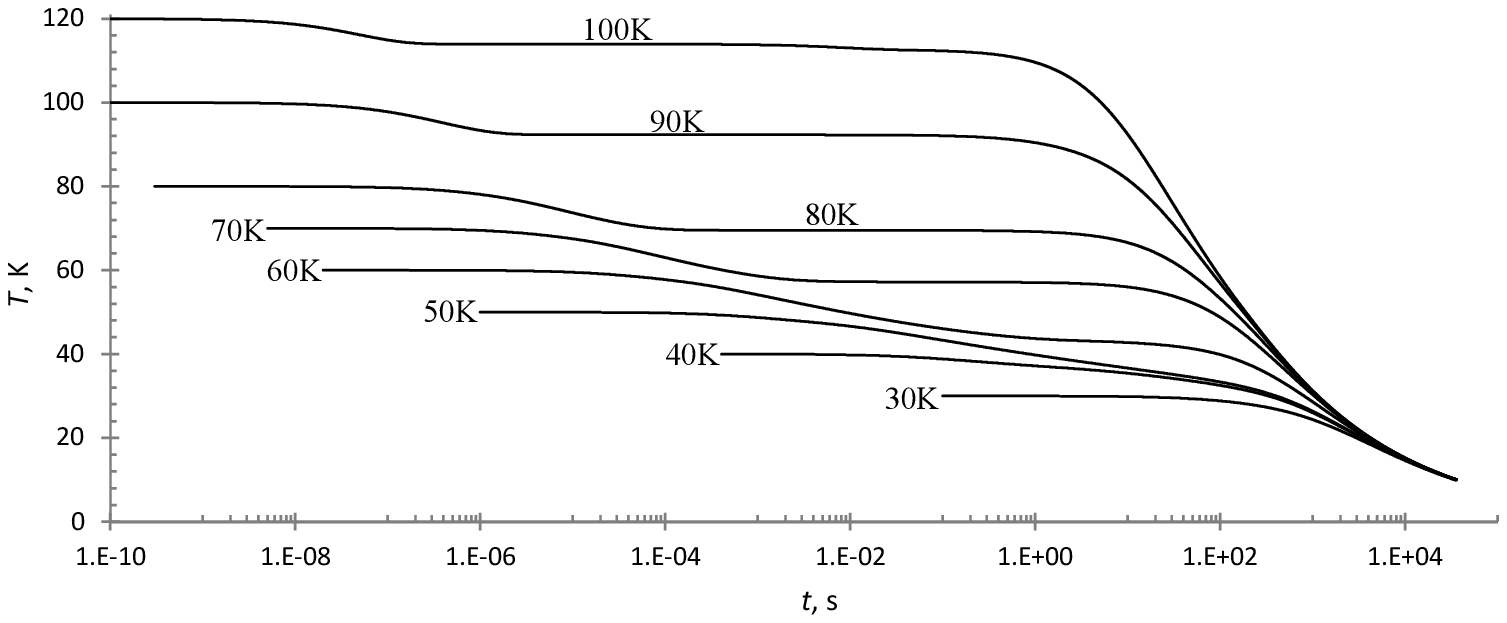}
		\vspace{-22cm}
   \caption{Temperature evolution for grain cooling simulations without diffusion and subsequent sublimation of bulk-ice molecules. The initial temperature $T_0$ is indicated for each curve.}
              \label{fig-ndT}
    \end{figure*}
The default configuration of the \textsc{Tcool} model includes the diffusion of bulk-ice molecules. Only diffusion toward the outer surface that eventually results in sublimation was considered, with data from \citet{Fayolle11}. Molecules deep in the icy mantle cannot diffuse and are trapped (see Eq.~(\ref{los3})). While there is evidence from temperature-programmed experiments that such bulk-ice diffusion and entrapment occur \citep{Oberg09,Oberg10,Fayolle11,Martin14,Tachibana17,Simon19}, the diffusion may be caused by molecules hopping on the surface of pores and cracks in ices, not actual bulk-ice molecule movement \citep{Lauck15,Cooke18}. No evidence has been found for such porosity of ices in the ISM \citep{Keane01,Pontoppidan03}. Thus, diffusion in the volume of an icy mantle on a heated grain might be possible only along channels filled with volatile molecules.

In order to investigate how efficient sublimation is in the simple case without bulk-ice diffusion, we performed nine simulations with $T_0$ in the range of 30 to 120\,K, for an $a=0.1\,\mu$m grain covered with 100\,MLs of ice. The Leger-Zhao heat capacity approach was employed.
Figure~\ref{fig-nd} shows that for initial peak temperatures $T_0$ up to about 50\,K, the simulations without diffusion produce a similar number of sublimated molecules to simulations with diffusion. For higher $T_0$, the number of surface molecules is insufficient to fully cool the grain, and diffusive sublimation becomes significant. The value of this threshold temperature can vary and is influenced by a number of parameters in the model: the assumed average size of icy molecules, grain size, ice thickness and composition, and grain heat capacity approach.

At $T_0=50$\,K in the present model, $N_{\rm ev.}\approx2.4\times10^6$ (both simulations), which, at 120\,K, grows to $2.7\times10^7$ for the simulation with diffusion and only $4.1\times10^6$ without diffusion. The latter number corresponds to 1.5--1.9\,MLs of ice. More than one surface ML can sublimate because sublimating molecules expose the surface beneath them, allowing desorption to continue until the surface is fully covered by non-sublimating species. In the no-diffusion model with $T_0\gtrsim 65$\,K, cooling is always dominated by radiation and the total number of desorbed molecules $N_{ev.}$ remains relatively constant. At $T_0=120$\,K, only 16\,\% of grain thermal energy is sublimated, while the rest is radiated. 

Given the dearth of volatile molecules on the surface, surface methane and carbon dioxide are sublimated for simulations with $T_0>50$ and $T_0>90$\,K, respectively. The number of sublimated methane molecules $N_{\rm ev.CH_4}$ reaches $\sim4\times10^4$ at $T_0=60$\,K and remains at this value for all simulations with higher $T_0$. Significant amounts of CO$_2$ ($N_{\rm ev.CO_2}>10^4$) are desorbed for $T_0>90$\,K, while at $T_0=120$\,K, $N_{\rm ev.CO_2}=2\times10^5$ (1\% of all CO$_2$ in the mantle). The removal of surface CO$_2$ exposes an additional part of the layer beneath, allowing slightly more other volatiles to be sublimated.

Interestingly, the number of desorbed molecules for high-$T_0$ simulations without diffusion is more comparable to those obtained with the original approach on CRD by \citet{Hasegawa93}. These authors assumed that for $T_0=70$\,K, grain cooling lasts only for the characteristic CO sublimation timescale, which is $\sim10^{-5}$\,s at this temperature, equaling $\sim10^6$ sublimated CO molecules. The \textsc{Tcool} model gives much higher CRD yields ($N_{\rm ev.CO}\approx5\times10^6$, cf. Fig.~8 of \I); however, if bulk-ice molecule diffusion does not occur in icy interstellar grains heated by CRs, the CRD yields are in the middle between these two values.

The temperature curves for simulations without diffusion, shown in Fig.~\ref{fig-ndT}, demonstrate the effects of the lack of surface volatiles. For simulations with $T_0\geq60$\,K, two distinct stages of temperature $T$ decrease can be distinguished. The first is caused by surface sublimation, while the second by radiative cooling, which sets in for integration times $t$ longer than about one second. The first $T$ decrease stage also illustrates the extent to which the grain can be cooled by surface sublimation alone. This can be compared to the cooling with the diffusive sublimation of bulk-ice molecules, which occurs via the continuous and overlapping sublimation of molecules with increasing desorption energy and depth in ice, resulting in a steady decrease of $T$, as discussed in \I\  (cf. Fig.~3 in that study).

\section{Summary}
\label{concl}

We have performed a series of simulations considering various aspects regarding the sublimation of molecules from interstellar grains heated above the ambient temperature. The main results are listed below.
\begin{itemize}
\item Heat capacity is an important parameter when grain initial temperature is in the vicinity of $T_0=40$\,K. High $C$ translates into a higher number of sublimated molecules $N_{\rm ev.}$. However, materials with lower heat capacity curves are more efficient at converting their heat energy content into the sublimation of molecules.
\item Cosmic-ray-induced desorption is most effective for medium-small grains with an approximate size of 0.02\,$\mu$m, probably supporting the conclusions by \citet{Iqbal18}.
\item A maximum of about six to seven CO molecules can be sublimated per electronvolt of energy deposited in icy interstellar grains.
\item Cosmic-ray-induced desorption of carbon dioxide occurs for small grains ($a<0.05\,\mu$m) that can be heated to high temperatures and do not have significant amounts of other volatiles.
\item The presence of H$_2$ molecules adsorbed onto the surfaces of icy grains reduces the sublimation of other volatile molecules for grains heated up to $T_0\approx30$\,K.
\item We simulated a case when the diffusion and subsequent sublimation of bulk-ice molecules does not occur in the icy mantles of interstellar grains. These simulations show a decreased desorption yield for $T_0>50$\,K, for all species, compared to simulations with such diffusion.
\end{itemize}
To summarize, we have elaborated on and clarified a number of questions regarding molecule sublimation from grains. The understanding acquired in this study will be essential in future astrochemical studies considering CRD or other processes involving a sudden heating and subsequent cooling of icy interstellar grains.

\begin{acknowledgements}
JK has been funded by ERDF postdoctoral grant No.~1.1.1.2/VIAA/I/16/194 ‘Chemical effects of cosmic ray induced heating of interstellar dust grains’. JRK has been funded by Latvian Science Council project No.~lzp-2018/1-0170 ‘Evolution of Organic Matter in the Regions of Star and Planet Formation (OMG)’. Both projects are being implemented in Ventspils University of Applied Sciences.
\end{acknowledgements}

   \bibliographystyle{aa}
   \bibliography{dzes2}

\begin{thebibliography}{35}
\expandafter\ifx\csname natexlab\endcsname\relax\def\natexlab#1{#1}\fi

\bibitem[{{Boogert} {et~al.}(2013){Boogert}, {Chiar}, {Knez}, {{\"O}berg},
  {Mundy}, {Pendleton}, {Tielens}, \& {van Dishoeck}}]{Boogert13}
{Boogert}, A.~C.~A., {Chiar}, J.~E., {Knez}, C., {et~al.} 2013, ApJ, 777, 73

\bibitem[{{Boogert} {et~al.}(2011){Boogert}, {Huard}, {Cook}, {Chiar}, {Knez},
  {Decin}, {Blake}, {Tielens}, \& {van Dishoeck}}]{Boogert11}
{Boogert}, A.~C.~A., {Huard}, T.~L., {Cook}, A.~M., {et~al.} 2011, ApJ, 729, 92

\bibitem[{{Bringa} \& {Johnson}(2004)}]{Bringa04}
{Bringa}, E.~M. \& {Johnson}, R.~E. 2004, \apj, 603, 159

\bibitem[{{Cooke} {et~al.}(2018){Cooke}, {{\"O}berg}, {Fayolle}, {Peeler}, \&
  {Bergner}}]{Cooke18}
{Cooke}, I.~R., {{\"O}berg}, K.~I., {Fayolle}, E.~C., {Peeler}, Z., \&
  {Bergner}, J.~B. 2018, \apj, 852, 75

\bibitem[{{Cuppen} {et~al.}(2006){Cuppen}, {Morata}, \& {Herbst}}]{Cuppen06}
{Cuppen}, H.~M., {Morata}, O., \& {Herbst}, E. 2006, MNRAS, 367, 1757

\bibitem[{{Draine} \& {Li}(2001)}]{Draine01}
{Draine}, B.~T. \& {Li}, A. 2001, \apj, 551, 807

\bibitem[{{Fayolle} {et~al.}(2011){Fayolle}, {{\"O}berg}, {Cuppen}, {Visser},
  \& {Linnartz}}]{Fayolle11}
{Fayolle}, E.~C., {{\"O}berg}, K.~I., {Cuppen}, H.~M., {Visser}, R., \&
  {Linnartz}, H. 2011, \aap, 529, A74

\bibitem[{{Hasegawa} \& {Herbst}(1993)}]{Hasegawa93}
{Hasegawa}, T.~I. \& {Herbst}, E. 1993, \mnras, 261, 83

\bibitem[{{Herbst} \& {Cuppen}(2006)}]{Herbst06}
{Herbst}, E. \& {Cuppen}, H.~M. 2006, Proceedings of the National Academy of
  Science, 103, 12257

\bibitem[{{Iqbal} \& {Wakelam}(2018)}]{Iqbal18}
{Iqbal}, W. \& {Wakelam}, V. 2018, Astronomy and Astrophysics, 615, A20

\bibitem[{{Kalv{\= a}ns}(2016)}]{K16}
{Kalv{\= a}ns}, J. 2016, \apjs, 224, 42

\bibitem[{{Kalv{\= a}ns}(2018)}]{K18}
{Kalv{\= a}ns}, J. 2018, \apjs, 239, 6

\bibitem[{{Kalv{\= a}ns} \& {Kalnin}(2019)}]{KK19}
{Kalv{\= a}ns}, J. \& {Kalnin}, J.~R. 2019, \mnras, 486, 2050

\bibitem[{{Kalv{\={a}}ns} \& {Kalnin}(2020)}]{KK20}
{Kalv{\={a}}ns}, J. \& {Kalnin}, J.~R. 2020, \aap, 633, A97 (Paper~I)

\bibitem[{{Kalv{\={a}}ns} \& {Shmeld}(2010)}]{KS10}
{Kalv{\={a}}ns}, J. \& {Shmeld}, I. 2010, \aap, 521, A37

\bibitem[{{Keane} {et~al.}(2001){Keane}, {Boogert}, {Tielens}, {Ehrenfreund},
  \& {Schutte}}]{Keane01}
{Keane}, J.~V., {Boogert}, A.~C.~A., {Tielens}, A.~G.~G.~M., {Ehrenfreund}, P.,
  \& {Schutte}, W.~A. 2001, \aap, 375, L43

\bibitem[{{Krumhansl} \& {Brooks}(1953)}]{Krumhansl53}
{Krumhansl}, J. \& {Brooks}, H. 1953, \jcp, 21, 1663

\bibitem[{{Lauck} {et~al.}(2015){Lauck}, {Karssemeijer}, {Shulenberger},
  {Rajappan}, {{\"O}berg}, \& {Cuppen}}]{Lauck15}
{Lauck}, T., {Karssemeijer}, L., {Shulenberger}, K., {et~al.} 2015, \apj, 801,
  118

\bibitem[{{Leger} {et~al.}(1985){Leger}, {Jura}, \& {Omont}}]{Leger85}
{Leger}, A., {Jura}, M., \& {Omont}, A. 1985, \aap, 144, 147

\bibitem[{{Mart{\'\i}n-Dom{\'e}nech} {et~al.}(2014){Mart{\'\i}n-Dom{\'e}nech},
  {Mu{\~n}oz Caro}, {Bueno}, \& {Goesmann}}]{Martin14}
{Mart{\'\i}n-Dom{\'e}nech}, R., {Mu{\~n}oz Caro}, G.~M., {Bueno}, J., \&
  {Goesmann}, F. 2014, \aap, 564, A8

\bibitem[{{{\"O}berg} {et~al.}(2011){{\"O}berg}, {Boogert}, {Pontoppidan}, {van
  den Broek}, {van Dishoeck}, {Bottinelli}, {Blake}, \& {Evans}}]{Oberg11}
{{\"O}berg}, K.~I., {Boogert}, A.~C.~A., {Pontoppidan}, K.~M., {et~al.} 2011,
  \apj, 740, 109

\bibitem[{{{\"O}berg} {et~al.}(2009){{\"O}berg}, {Fayolle}, {Cuppen}, {van
  Dishoeck}, \& {Linnartz}}]{Oberg09}
{{\"O}berg}, K.~I., {Fayolle}, E.~C., {Cuppen}, H.~M., {van Dishoeck}, E.~F.,
  \& {Linnartz}, H. 2009, \aap, 505, 183

\bibitem[{{{\"O}berg} {et~al.}(2010){{\"O}berg}, {van Dishoeck}, {Linnartz}, \&
  {Andersson}}]{Oberg10}
{{\"O}berg}, K.~I., {van Dishoeck}, E.~F., {Linnartz}, H., \& {Andersson}, S.
  2010, \apj, 718, 832

\bibitem[{{Pauly} \& {Garrod}(2016)}]{Pauly16}
{Pauly}, T. \& {Garrod}, R.~T. 2016, \apj, 817, 146

\bibitem[{{Pontoppidan} {et~al.}(2003){Pontoppidan}, {Fraser}, {Dartois},
  {Thi}, {van Dishoeck}, {Boogert}, {d'Hendecourt}, {Tielens}, \&
  {Bisschop}}]{Pontoppidan03}
{Pontoppidan}, K.~M., {Fraser}, H.~J., {Dartois}, E., {et~al.} 2003, \aap, 408,
  981

\bibitem[{{Rawlings} {et~al.}(2013){Rawlings}, {Williams}, {Viti},
  {Cecchi-Pestellini}, \& {Duley}}]{Rawlings13}
{Rawlings}, J.~M.~C., {Williams}, D.~A., {Viti}, S., {Cecchi-Pestellini}, C.,
  \& {Duley}, W.~W. 2013, \mnras, 430, 264

\bibitem[{{Shen} {et~al.}(2004){Shen}, {Greenberg}, {Schutte}, \& {van
  Dishoeck}}]{Shen04}
{Shen}, C.~J., {Greenberg}, J.~M., {Schutte}, W.~A., \& {van Dishoeck}, E.~F.
  2004, \aa, 415, 203

\bibitem[{{Simon} {et~al.}(2019){Simon}, {{\"O}berg}, {Rajappan}, \&
  {Maksiutenko}}]{Simon19}
{Simon}, A., {{\"O}berg}, K.~I., {Rajappan}, M., \& {Maksiutenko}, P. 2019,
  \apj, 883, 21

\bibitem[{{Tachibana} {et~al.}(2017){Tachibana}, {Kouchi}, {Hama}, {Oba},
  {Piani}, {Sugawara}, {Endo}, {Hidaka}, {Kimura}, {Murata}, {Yurimoto}, \&
  {Watanabe}}]{Tachibana17}
{Tachibana}, S., {Kouchi}, A., {Hama}, T., {et~al.} 2017, Science Advances, 3,
  eaao2538

\bibitem[{{Wei} {et~al.}(2005){Wei}, {Wang}, \& {Wang}}]{Wei05}
{Wei}, Y.~X., {Wang}, R.~J., \& {Wang}, W.~H. 2005, \prb, 72, 012203

\bibitem[{{Whittet} {et~al.}(2011){Whittet}, {Cook}, {Herbst}, {Chiar}, \&
  {Shenoy}}]{Whittet11}
{Whittet}, D.~C.~B., {Cook}, A.~M., {Herbst}, E., {Chiar}, J.~E., \& {Shenoy},
  S.~S. 2011, ApJ, 742, 28

\bibitem[{{Woon}(2004)}]{Woon04}
{Woon}, D.~E. 2004, Advances in Space Research, 33, 44

\bibitem[{{Xie} {et~al.}(2018){Xie}, {Ho}, {Li}, \& {Shangguan}}]{Xie18}
{Xie}, Y., {Ho}, L.~C., {Li}, A., \& {Shangguan}, J. 2018, \apj, 867, 91

\bibitem[{{Zhao} {et~al.}(2018){Zhao}, {Caselli}, \& {Li}}]{Zhao18}
{Zhao}, B., {Caselli}, P., \& {Li}, Z.-Y. 2018, \mnras, 478, 2723

\bibitem[{{Ziegler} {et~al.}(2010){Ziegler}, {Ziegler}, \&
  {Biersack}}]{Ziegler10}
{Ziegler}, J.~F., {Ziegler}, M.~D., \& {Biersack}, J.~P. 2010, Nuclear
  Instruments and Methods in Physics Research B, 268, 1818

\end{thebibliography}

\end{document}